# Bacterial floc mediated rapid streamer formation in creeping flows

Mahtab Hassanpourfard[a], Zahra Nikakhtari[b], Ranajay Ghosh[c], Siddhartha Das[d], Thomas Thundat[a], Yang Liu[e], Aloke Kumar[b]*

[a]Department of Chemical and Materials Engineering, University of Alberta, Edmonton, Canada

[b]Department of Mechanical Engineering, University of Alberta, Edmonton, Canada

[c]Department of Mechanical and Industrial Engineering, Northeastern University, Boston MA 02115, USA

[d] Department of Mechanical Engineering, University of Maryland, College Park, MD 20742, USA

[e]Department of Civil and Environmental Engineering, University of Alberta, Edmonton, Canada

**Abstract:** One of the central puzzles concerning the interaction of low Reynolds number ($Re \ll 1$) fluid transport with bacterial biomass is the formation of filamentous structures called streamers. In this manuscript, we report our discovery of a new kind of low $Re$ bacterial streamers, which appear from pre-formed bacterial flocs. In sharp contrast to the biofilm-mediated streamers, these streamers form over extremely small timescales (less than a second). Our experiments, carried out in a microchannel with micropillars rely on fluorescence microscopy techniques to illustrate that floc-mediated streamers form when a freely-moving floc adheres to the micropillar wall and gets rapidly sheared by the background flow. We also show that at their inception the deformation of the flocs is dominated by recoverable large strains indicating significant elasticity. These strains subsequently increase tremendously to produce filamentous streamers. Interestingly, we find that these fully formed streamers are not static structures and show viscous response at time scales larger than their formation time scales. Finally we show that such novel streamer formation can lead to rapid clogging of microfluidic devices.

# Introduction:

In their natural state bacteria can be found in either disparate planktonic forms or living in tight knit communities such as flocs, mats, pellicles or biofilms [1-3]. The latter, aggregative modes of bacterial growth are characterized by cells embedded in a matrix, usually of self-produced extracellular polymeric substances (EPS) composed of long-chain biomolecules such as polysaccharides, nucleic acids and lipids [1,4-7]. This composite soft matter, consisting of bacteria and EPS, has been attracting intense scrutiny due to a complex interplay between material behavior and the underlying life processes brought about by large deformation even at very low Reynold's number ($Re$); for example, in the case of filamentous bacterial streamers generating from bacterial biofilms [8-11].

Streamers are so named due to their distinguishing filamentous morphology. They have been reported in systems with sustained hydrodynamic flows [9]. These slender bacterial aggregates are typically tethered at one or both ends to solid surfaces, while the rest of the structure is suspended in a liquid environment. Bacterial streamers have been observed to form both in high [12,13], and in low Reynolds number conditions ($Re<<1$) [10,11,14-16]. Streamer formation in low $Re$ transport is of significant technological and biomedical interest due to relevance to a wide variety of critical operating scenarios including clogging of biomedical devices such as heart stents, catheters, porous media and water filtration systems [8,11,17]. Rusconi et al. [10], while studying the effect of curved channel geometries using a microfluidic device, found that *Pseudomonas aeruginosa* formed streamers in the curved sections of the microchannels. Drescher et al. [8] later showed that streamer formation in microfluidic devices with curved sections can lead to catastrophic clogging. Valiei et al. [11] used a microfluidic device with micropillars, and studied biofilm formation by *Pseudomonas fluorescens*. They found that, in a certain flow regimes, the bacteria formed extensive streamers resulting in a web-like network between the different pillars. A commonality between the reports by Valiei et al. [11], Rusconi et al. [10] and Drescher et al. [8] is that the streamers appeared far later than the biofilms, and the corresponding streamer formation time-scale, $t_s$, was of the order of hours from the beginning of the flow. Das and Kumar [18] have recently proposed that in such instances, where the streamer formation time-scale far exceeded the relaxation time-scale of biofilms, streamers appeared from a highly viscous state of the intrinsically viscoelastic biofilms. In contrast to these studies, some other experiments conducted under apparently similar creeping flow conditions reported much smaller $t_s$ values ($t_s \sim$ minutes)[15,16] (see Table 1). Kim et al. [19] have reported even smaller streamer formation time scales for the bacterium *Staphylococcus aureus*, though this was achieved by first coating the channel walls of the microfluidic device with human plasma. Such large variation in streamer formation time scale might indicate different physical mechanisms that govern the streamer formation process. Streamers forming at very large time scales ($t_s \sim hrs$) have typically been reported in systems where formation of a biofilm occurs prior to streamer formation; referred herein as biofilm-mediated streamer formation [8,10,11]. To the best of our knowledge, a proper quantitative evaluation of small time scale streamer formation is yet to be reported. Furthermore, much of the literature on streamer formation in low Reynolds number conditions is relatively recent in the context of literature on biofilms and the physical basis of streamer formation remains an active area of research[9,20,21].

In this study, we report our discovery of a new kind of bacterial streamer formation – these streamers do not appear from biofilms, rather they appear due to flow-induced deformation of the pre-formed bacterial flocs. We conduct our experiments in a microfluidic device composed of micropillars and containing solution laden with flocs of the bacterium *Pseudomonas flourescens*. We are able to optically probe the inception process of the streamers by embedding the bacterial flocs with 200 nm red fluorescent polystyrene beads that serve as tracers. We discover that fluid flow first advects the flocs some of which then get attached to the micropillars; subsequently, the hydrodynamic shear forces deform them into filamentous streamers. Interestingly, the formation timescale of this floc-mediated streamer formation is less than a second, which is in sharp contrast to the much larger timescale witnessed for biofilm-mediated streamers. Next, we find that the streamers are not purely elastic structures since we observed perceptible viscous behavior at time scales larger than its formation time scale but still far from clogging regime. Our methodology of using nanoscale fluorescent tracers, allows for the first time a direct quantification of the evolution of the streamer morphology. This gives us valuable clues about the fundamental mechanism of floc-mediated streamer formation in contrast to the much more extensively studied biofilm-mediated streamer formation, which still remains a highly contentious problem.

## Results

**Initiation of floc-mediated streamers**

Our microfluidic device (Fig. 1a,b) consisted of a sequence of polydimethylsiloxane (PDMS) micropillars in a periodic staggered grid pattern. The micropillars had a diameter ($d$) of 50 μm and spaced 75 μm apart ($l$). The fluid flow rate ($Q$) was maintained at levels such that the resultant flow in our device was in the creeping flow regime (Reynolds number, $Re$, was $O(10^{-3})$). Numerical simulations provide the velocity profile inside the channel under these conditions. Fig 1c depicts the non-dimensionalized (with respect to $U = Q/(W \times h)$ where $W$ & $h$ are channel width and height respectively) contour plot of magnitude of velocity and also streamlines (inset) in the device.

We studied the behavior of the wild type (WT) strain of *Pseudomonas flourescens*, a bacteria that plays a vital role in maintaining plant physiology [22]. The genetically modified WT expressed green fluorescent protein (GFP) constitutively and hence was green fluorescent. Preformed biomass of *P. flourescens* in the form of bacterial flocs was utilized in our study (Fig 2a). Bacterial flocs are EPS encapsulated aggregates of the bacteria that are dispersed in a liquid phase. These flocs were imaged first in quiescent media and their equivalent diameter was measured. For our system we observed a wide variation in the equivalent diameter. Their quantification was done through a relative frequency histogram, which shows that the mode for these flocs occurred at approximately 22 μm (Fig. 2b). These flocs were further mixed with 200 nm red fluorescent amine coated polystyrene (PS) particles and the mixture was allowed to flow through the microfluidic device. Two color imaging was performed for the system as the bacteria were green fluorescent and PS particles were red fluorescent. The red fluorescent beads were embedded in the EPS matrix of the flocs and thus provided clear visualization of the dynamics of these flocs (Fig. 2c & d). Such two-color visualization helps us overcome the difficulty in visualizing EPS, which is almost transparent under brightfield illumination.

The solution containing planktonic bacteria and bacterial flocs was flown through the device for several minutes and it was observed that streamer like structures formed within a few minutes of the initiation of the experiment (Fig. 3 & Supplementary Video 1). The bacterial flocs could be seen to attach to the micro-pillar posts and then deformed by fluid shear. A dashed ellipse marks the location of this event. Similarly, in the right hand side, another floc undergoes a similar process. It is interesting to note that the size distribution of flocs measured in quiescent media (Fig. 2a) is not reflected in the size of flocs that were observed to flow past the pillars (see Fig S1). For instance, although the mode of floc sizes was similar to the pillar diameter, such flocs were not observed in the flow past these pillars (see Supplementary information Fig. S1).

**Timescale and mechanism of floc-mediated streamer formation**

The central result of our study is that the time-scale of formation of these floc-mediated streamers ($t_s$) is very small (Fig. 3 & Supplementary Video 1). A close examination of Supplementary video 1 reveals that the floc in the right hand side of Fig 3 undergoes large deformation at $t_s < 1$ sec. For instance, in the left hand side of Fig. 3, a floc approximately 3-3.5 μm in diameter undergoes very large deformation to form a streamer like structure. In Supplementary Video 1, a constant volume flow rate (or a constant $U$) is enforced by the syringe pump; however in the initial period when all pillars have not been wetted, $U$ can have considerable oscillatory component in time[23]. Let us denote the time for this initial wetting as $t_{wetting}$, which is observed to be about ($\sim 95$ s) and taken to be the time when $U$ becomes constant indicating the onset of completion of wetting. Thus at $t > t_{wetting}$, we assume that wetting of all pillars is complete, and $U$ becomes constant. For $t < t_{wetting}$, i.e. when $U$ has an oscillatory component in time, the response of the flocs to a temporally varying fluid shear at a spatial location can be seen clearly (Fig 4 and Supplementary Video 1). Probing further, we track a set of closely placed particle couplets and measure the ratio of their separation along the streamer to their initial separation. This stretch ratio (λ) contains useful information regarding the material behavior of the streamer. To this end, a floc is chosen where embedded PS beads act as tracers and allow us to identify two closely situated points, $\alpha$ and $\beta$, and then these are tracked as a function of time (Fig. 4a). The point $\alpha$ is largely immobile due to its adhesion to the cylinder wall, while $\beta$ is displaced by the fluid shear forces. As the fluid shear force scales with velocity $\tau \sim \mu \frac{U}{L}$, a time-periodic $U$ results in a time-periodic $\tau$. The distance between the two points in their initial (reference state) is denoted by $d\mathbf{X}$ and in the current state by $d\mathbf{x}$. The axial stretch ratio defined as $\lambda(t) = \frac{|d\mathbf{x}|}{|d\mathbf{X}|}$ is plotted with respect to time in Figure 4b. As expected, Figure 4b indicates that, at $t < t_{wetting}$, $\lambda(t)$ oscillates between unity and a maximum value of approximately 3 (i.e. 200% engineering strain). This initial recoverable strain clearly indicates an elastic component of the streamer material. As the flow velocity increases, the streamer stretching begins to increase until, at the advent of steady flow, the floc is stretched into the slender geometry characteristic of streamer. After the initial wetting period ($t > 95$ s), the streamer becomes attached between two successive pillar walls indicating stretch of the order of seven (Fig 4b). If a complete loss of material strength is assumed at this deformation, a lower limit of streamer formation time can be estimated from the background fluid velocity profile obtained from the simulations (Fig 1c). Assuming an average transport speed of approximately $U = 13 \times 10^{-5}$ m/s, and transport distance as the inter-pillar separation length, of $l = 75 \times 10^{-6}$ m, we get $t_s \sim \frac{l}{U}$ which comes to be approximately of the order of $t_s$ is $O(10^{-1})$ s. This time scale agrees well with our experiments. Furthermore, from the numerical simulations, the role of shear deformation in streamer formation is strongly suggested as well (see Fig 5), since most of the streamers originate from regions corresponding to maximum shear stress.

Streamers eventually lead to catastrophic clogging of the device, such as that observed by Drescher et al.[8] Interestingly however, the time period spanning the advent of streamer formation to final clogging is not marked by a sudden transition if closer look at the streamer behavior is taken. In this context, on an experimental time scale greater than the streamer formation time-scale ($t > t_s$), but still far from clogging related change of overall velocity profile and streamer shape, there is a perceptible viscous component indicated by a creeping response of a material point of the streamer material (Supplementary Video 2). Quantification of this response is made possible by evaluation of the temporal response of the principal velocity gradient (Fig. 6) (see Materials and Methods). The plot shows that velocity gradient at a point remains approximately constant with time while the background velocity which scales with the loading shear stress is assumed to be approximately constant in the time window of observation indicating a steady creep regime under these flow conditions.

**Clogging action of the floc-mediated streamers**

At a time-scale much greater than the streamer formation time-scale $(t \gg t_s)$, a very different picture emerges. Streamer formation is a dynamic phenomenon where the streamer grows in width as it accrues additional mass from its surrounding fluid. Figure 7a shows that that in approximately one hour after the beginning of the experiment a large part of the device is covered in biomass. This can be quantified by measuring the surface coverage by the biofilm as a function of time. Figure 7b shows that very quickly about 50% of the device is covered by the bacterial film. Thus clogging in this device can not only be catastrophic [8] but also take place at a rapid clogging rate. The exponential increase in surface coverage can be explained by previously developed models for streamer growth as explained by Dreshcer et al. [8] and later corrected by Das and Kumar [18].

**Discussion**

Here, using seeded particles, which allowed very precise quantification of the deformation of the biomass structures we clearly demonstrate that pre-formed biomass, in the form of bacterial flocs, can lead to streamer formation through the process of large deformations, even when fluid flow in a system lies in the creeping flow regime. This is the first experimental observation, where the formation of a streamer has been demonstrated and we have shown that pre-formed biomass can lead to very rapid streamer formation time scales. Furthermore, particle tracking enabled us to conclude that the material behavior of the streamer can have both significant elastic as well as viscous component making them highly dynamic mechanical structures even in creeping flows. Finally, streamers cause not only catastrophic, but very rapid clogging of devices.

The exact material constitution of the bacterial communities is of great interest for a range of applications [9]. However, traditional material characterization techniques are impractical for *in situ* applications thus making the characterization of this system far more challenging. However, important material information is obtained by scrutinizing the response of the streamer due to fluidic loading by the background flow. The temporal behavior of the axial strain as quantified by $\lambda(t)$, offers important insights into the mechanics of streamer formation. An interesting aspect of the streamer formation is the ability of the biomass to remain elastic under relatively large stretch ratios (engineering strain ~200%). This is typical of elastomeric materials. Such large strain elasticity is typically attributed to a molecular level 'chain stretching' [24]. However, for the current material, it may also include straining of a more complex intermediate hierarchical microstructure well known in biofilm morphological literature[1,25]. This study would thus serve as an important motivation for such future extensions. As the flow develops fully, the shear stress on the incipient streamer increases significantly. This results in a significant increment of the shear stress causing the incipient structure to undergo even larger deformation after which the streamer extends between adjacent pillars and streamers are formed, see Fig 3. The strain regime corresponding to this highly extended state ($\lambda \sim 7$), typically indicate substantial inelastic behavior. This is confirmed by observing a material point couple which move slowly through the streamer even when the background flow is constant (Fig. 5). We assume that the measured velocity gradient at these deformations is almost entirely inelastic in nature.

Biofilm streamer formation remains an exciting frontier with several open ended unanswered questions. Here, we clearly demonstrate that pre-formed biomass, in the form of bacterial flocs, can lead to streamer formation through large deformations, even when fluid flow in a system lies in the creeping flow regime. This is the first direct experimental observation, where the formation of a streamer and their behavior in the intermediate time scale with respect to clogging has been demonstrated by using seeding particles for very precise quantification of the biomass structures. The key results of this work include demonstration of very rapid streamer formation and subsequent creep response of fully formed streamers. Finally, we show that streamer formation lead to rapid clogging of the device.

## Material and Methods:

**Microchip fabrication:**

A 4" silicon master mold was prepared by following the conventional photolithography process from the designed pattern. Then, by applying soft lithography processess and using the prepared master mold the final device was fabricated from polydimethylsiloxane (PDMS, Sylgard 184, Dow Corning, NY, USA). Next, by exposing the cover slip and the PDMS stamp to oxygen plasma for 30 seconds, the PDMS stamp and the cover slip were bonded together to prepare the microchip. At the end, the microchip was annealed at 70 °C for 10 min for attaining better sealing. The protocol is described in detail by Hassanpourfard et al. [14].

**Bacterial culture:**

In this experiment, colonies of *Pseudomonas fluorescens* CHA0 (wild type) were grown on Luria-Bertani (LB) agar plate at 30 °C overnight. Next, one colony from an agar plate was inoculated into LB broth medium. The strain is green fluorescent as they express green fluorescent protein (GFP) constitutively. To stimulate floc formation in the bacterial solution, the bacteria solution was kept in a shaker incubator (New Brunswick Scientific Co., NJ) at 30 °C and 150 rpm for about 48 h. Longer incubation duration leads to nutrient depletion and subsequent floc formation [26].

**Microscopy:**

The microfluidic chip was placed on the stage of an inverted optical microscope (Nikon Eclipse Ti). The bacteria solution was injected continuously into the microchip by using a syringe pump (Harvard Apparatus, MA, USA). The temperature of microchip was set at 30 °C by the aid of an on-stage microscope incubator (Pathologic Devices, Inc., MD, USA). Particle tracking was performed using the object tracking module in Nikon NIS-Element AR software interface. Effective surface area calculations were also performed using the same software. Error for stretch ratio calculation was estimated to be approximately 8%. Surface coverage percentage was calculated by defining a threshold value for green color intensity. If the green color intensity of a pixel was above the threshold value then corresponding pixel was counted. Next, we calculated the surface coverage percentage by dividing the counted pixels to the total number of pixels in the image (except area covered by the pillars) and multiplying by 100. This procedure is the same as that outlined in a previous publication by Kumar et al. [27].

**Computational fluid mechanical (numerical) simulations**

We performed 2-D fluid mechanical simulations using the commercial package Comsol Multiphysics® to simulate the device flow. In this simulation, we assumed that the fluid going through the channel has the same properties as water at 30°C and pressure. To describe the fluid flow in the channel the incompressible Navier-Stokes and continuity equations were used. The inlet velocity was calculated according to the flow rate (here 15 µl/h). The no-slip and the no penetration boundary condition were imposed on the walls and a constant atmospheric pressure was imposed at the channel outlet, respectively. Mesh density was increased until no mesh dependency was observed in the solution. Velocity is non-dimensionalized with respect to the velocity scale, $U = Q/(W \times h)$, where $Q$ is the volume flow rate (imposed by the syringe pump).

**Calculation Velocity Gradient**

We tracked a particle **P** which moves in a fixed Eulerian grid from $P_0(x_0, y_0)$ to $P_1(x_1, y_1)$ in time $\delta t_1$ and then further moves to $P_2(x_2, y_2)$ in another time $\delta t_2$.

The normalized velocity gradient at $P_0$ is then given by forward-difference discretization:

$$[L] \approx \frac{l}{U \, \delta t_1 \delta t_2} \begin{pmatrix} \frac{x_2 \delta t_1 - x_1(\delta t_1 + \delta t_2) + x_0 \delta t_2}{x_1 - x_0} & \frac{x_2 \delta t_1 - x_1(\delta t_1 + \delta t_2) + x_0 \delta t_2}{y_1 - y_0} \\ \frac{y_2 \delta t_1 - y_1(\delta t_1 + \delta t_2) + y_0 \delta t_2}{x_1 - x_0} & \frac{y_2 \delta t_1 - y_1(\delta t_1 + \delta t_2) + y_0 \delta t_2}{y_1 - y_0} \end{pmatrix} \qquad (1)$$

Two eigenvalues of $[L]$ corresponding to two principal eigenvectors were found. One of the eigenvectors was found to be nearly aligned with the orientation of the streamer and the eigenvalue corresponding to the other eigenvector was vanishingly small as expected. Thus for the purpose of this work, we denote the larger eigenvalue (which is aligned with the streamer orientation) of the velocity gradient tensor as the principal velocity gradient, $L_1$. Note that this principal velocity gradient component is assumed to be mostly inelastic in the time frame considered with elastic and rotational components assumed negligible.

**Tables:**

**Table 1:** Time scale of streamer formation from different experiments

| Bacteria | Streamer formation time scale (hr) | Comment | Ref. |
|---|---|---|---|
| *Pseudomonas aeruginosa* | 6-7 | | Rusconi et al. [10] |
| *Pseudomonas aeruginosa* | 18 | | Rusconi et al. [28] |
| *Pseudomonas fluorescens* | few hours | for certain flow rates (8-12-20 ul/h) | Valiei et al. [11] |
| *Pseudomonas aeruginosa* | 50 | | Drescher et al. [8] |
| *Staphylococcus epidermis* | 6 | | Weaver et al. [29] |
| *E. Coli* | 0.5 | | Yazdi and Ardekani [16] |
| **ature *Pseudomonas fluorescens*** | **$\sim 10^{-4}$ (i.e. a few seconds)** | | **Present work** |

**Figures:**

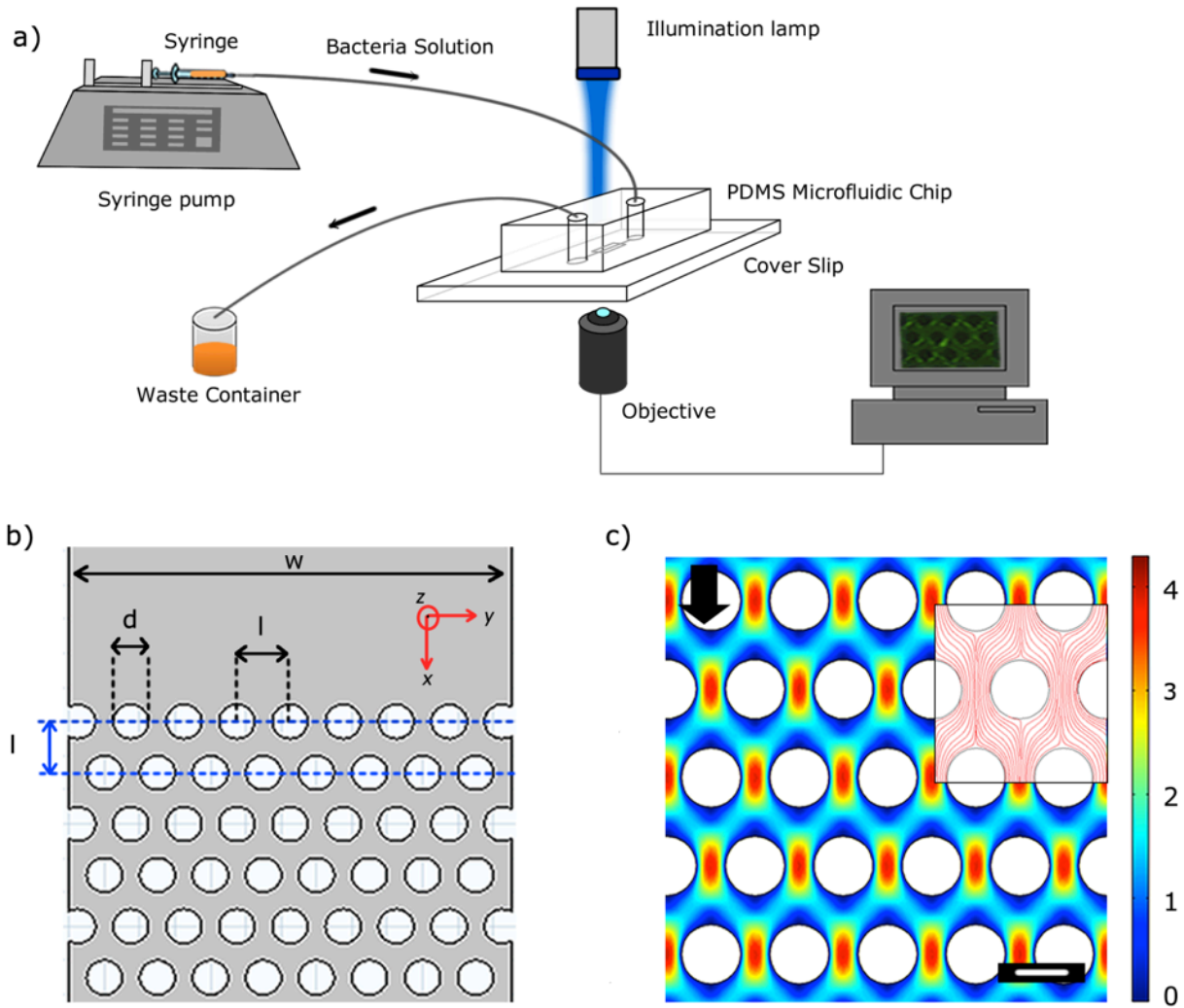

**Figure1**: a) A schematic of experimental set-up under pressure driven flow with constant volume flow rate ($Q$). b) layout of staggered pattern porous media. Width ($W$) of porous zone is 625 μm. The distance between the center of pillars ($l$) and 2 rows of consecutive pillars is 75 μm. The diameter of the pillars and the height of the device both are 50 μm. c) Computational fluid mechanical simulations demonstrating the non-dimensionalized velocity contour of the flow in the porous section of the microchannel. In our device $Q = 15\,\mu l/hr$ corresponds to $U = 13\times10^{-5}$ m/s. The scale bar is 50 μm. (Inset) Streamlines for the same flow condition.

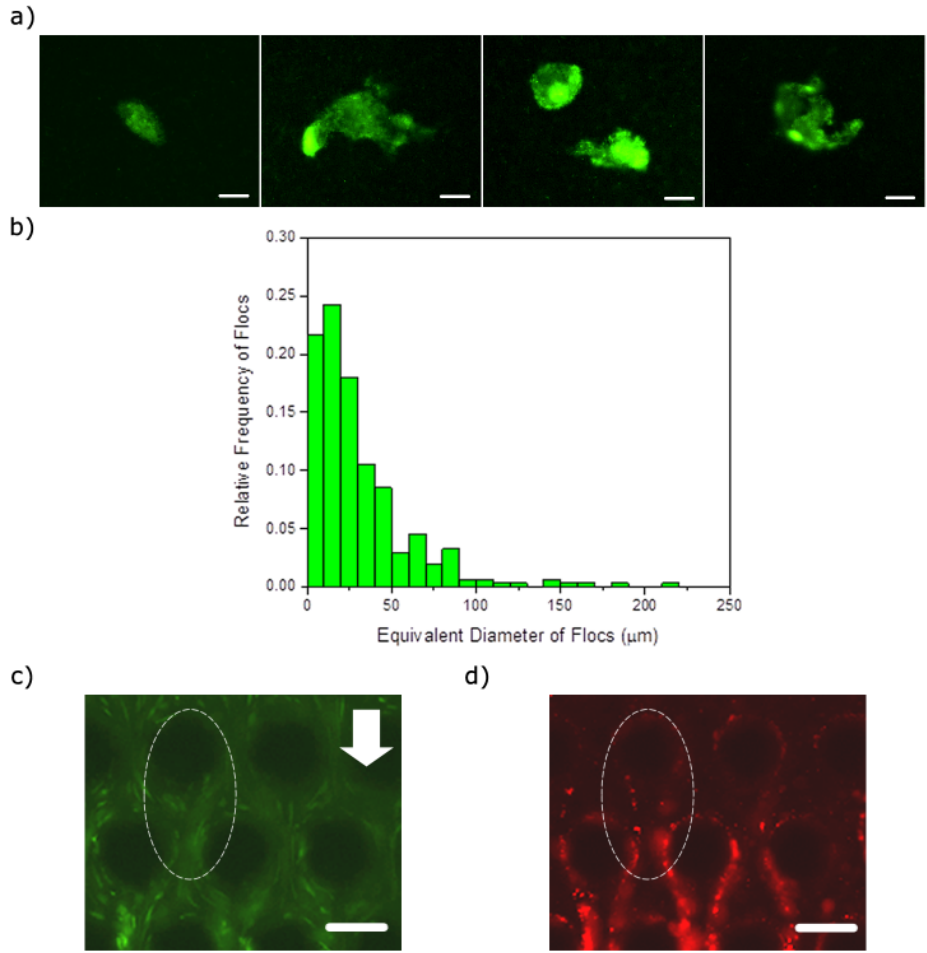

**Figure 2**: a) Flocs of *P. fluorescens* bacteria after incubation at 30 °C overnight. The scale bars are 25 μm. b) Relative frequency histogram of the flocs. The *x*-axis is equivalent diameter of the flocs in the reservoir. The median and mode for this relative frequency histogram are 21.28 and 22.23 μm, respectively. c) and d) Green and red fluorescent images, respectively, of the microchannel after injecting the bacteria with 200 nm fluorescent red polystyrene beads particles into it. The images were taken at the same time and place that was approximately at the middle of the channel height ($z$ =25 μm). The red fluorescent particles clearly seed and enable visualization of the streamer. Note the regions demarcated by dashed ellipses where bacteria (green) are not significant, but the streamer itself is easy visualized due to red particles seeding the EPS network. The scale bars are 50 μm.

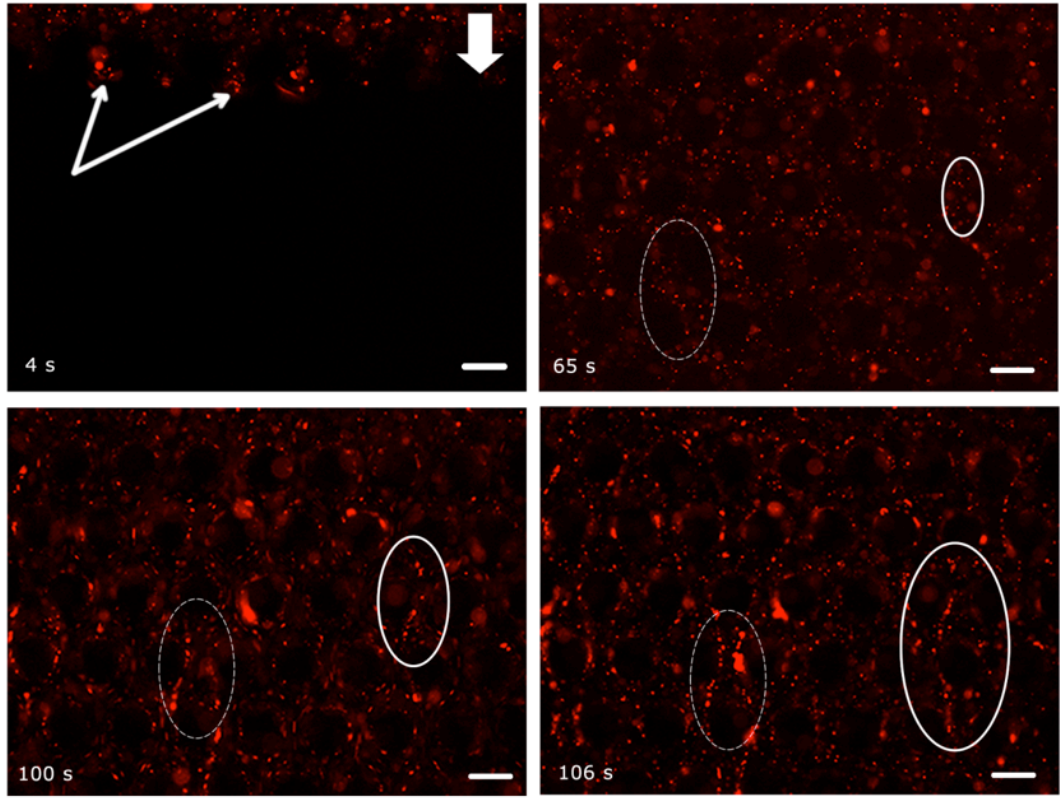

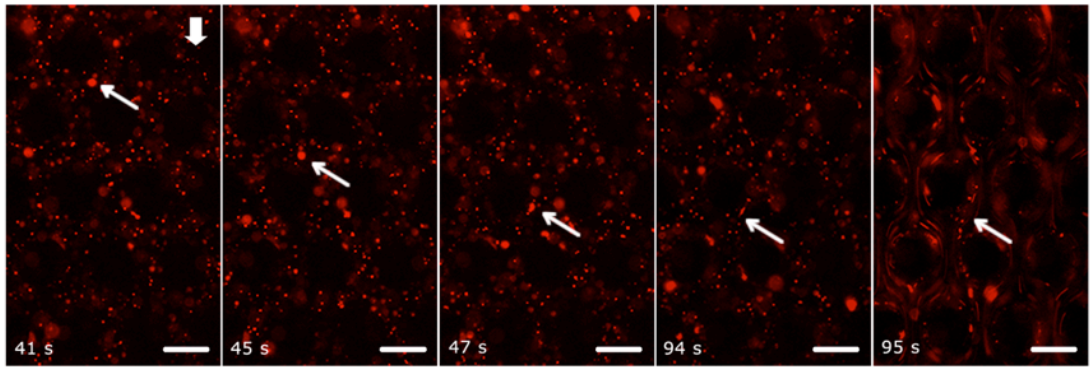

**Figure 3**: Scale-bars are 50 μm. (a) Rapid streamer formation in a short time scale (a few seconds). The scale bars are 50 μm. See accompanying video. The images were taken approximately at the middle of the channel height ($z=25$ μm). In the top-left image, the arrows demarcate the advancing fluid meniscus. The ellipses demarcate two regions where streamers form. (b) An arrow demarcates a floc, which is first advected through the channel and then is attached to a micropillar wall at $t=47$ s and finally at 95 s a streamer is formed.

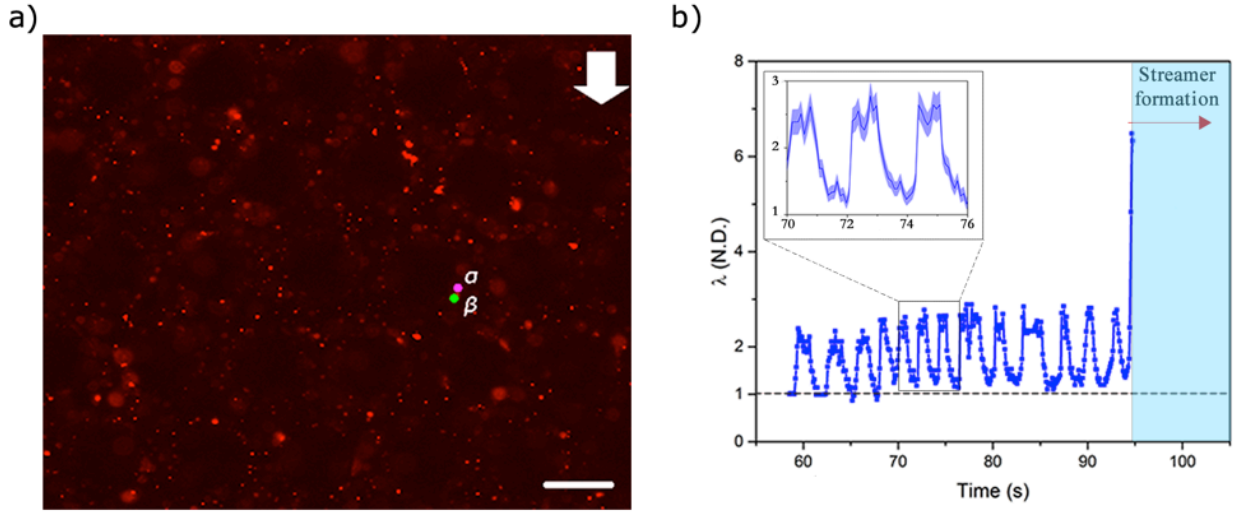

**Figure 4**: (a) Two points $\alpha$ and $\beta$ are tracked as the fluid velocity fluctuates in the channel. These two points would later form a streamer. The scale bar is 50 μm. (b) Stretch ratio of the two points as a function of time during the initial filling period of the channel. Streamer formation region (right) also coincides with the onset of steady flow. Black dashed line depicts λ=1 (Inset) Stretch ratio for a smaller time segment. The colored envelope represents estimated error.

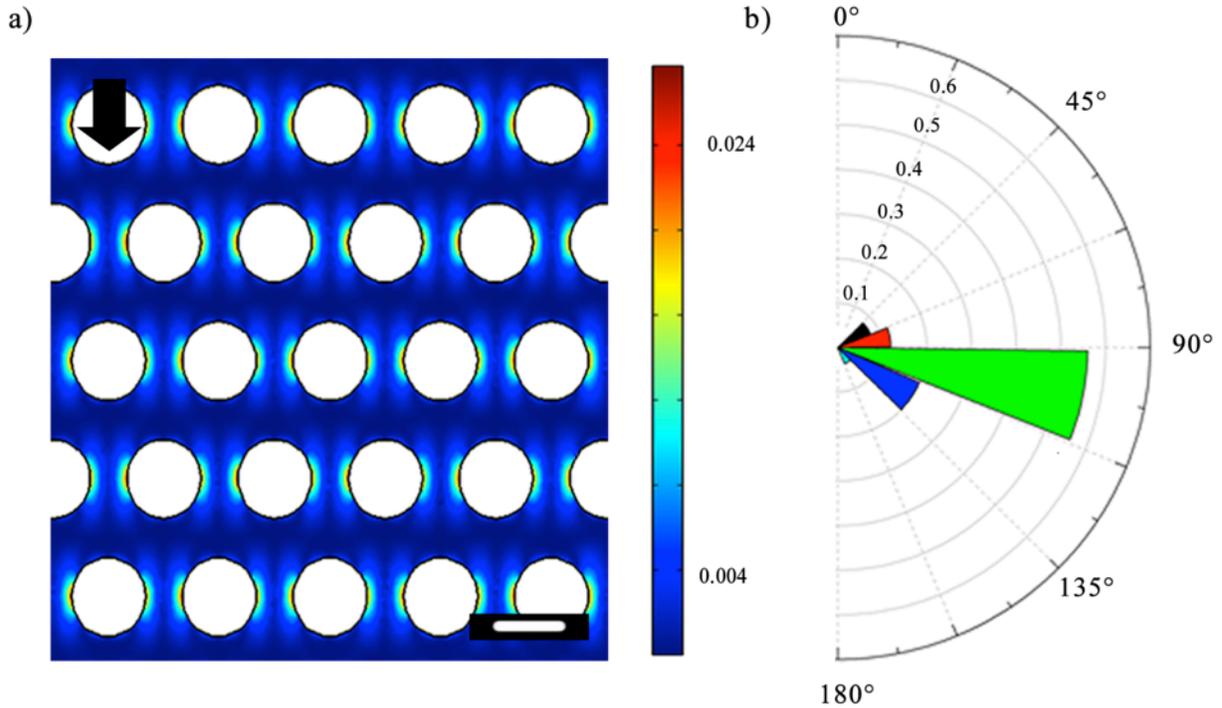

**Figure 5:** (a) Simulation results for the contour of the magnitude of the dimensionless shear stress $\bar{\tau} = \frac{d}{U} \mu \frac{1}{2}\left(\nabla u + (\nabla u)^T\right)$, $u$ being the velocity field (b) Polar frequency histogram of where flocs attach

on the pillars. Half of one pillar is considered. 0° represents upstream stagnation point and 180° represents downstream stagnation point.

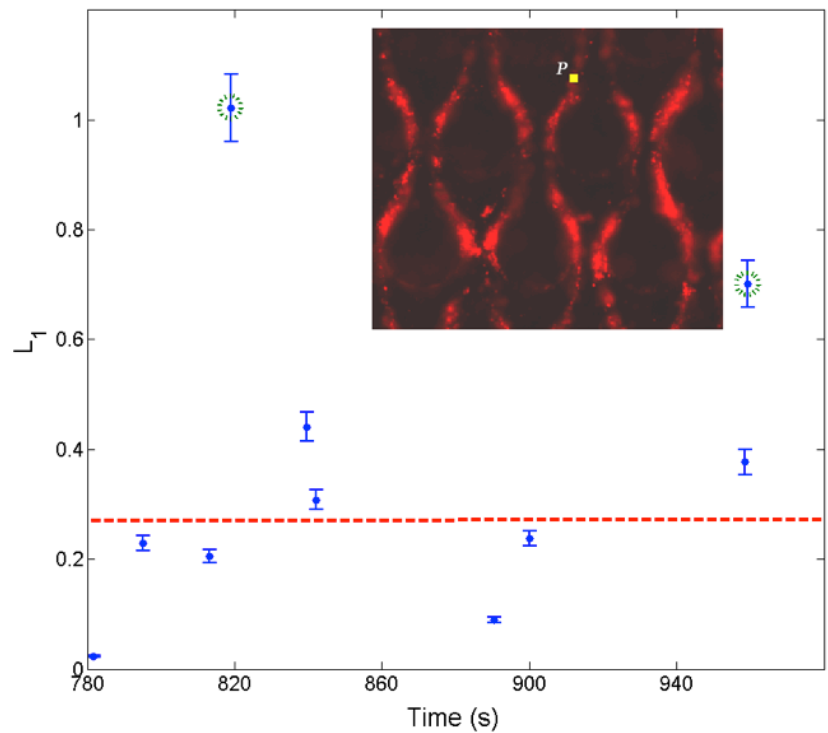

**Figure 6**: Temporal variation of the non-dimensional principal velocity gradient ($L_1$) of a fully formed streamer. The experimental obtained values points (blue dots) show an approximately constant trend (dashed red line); two points demarcated through dashed green ellipses were neglected as outliers (Inset) A point *P* was tracked on a fully formed streamer for calculation of the velocity gradient.

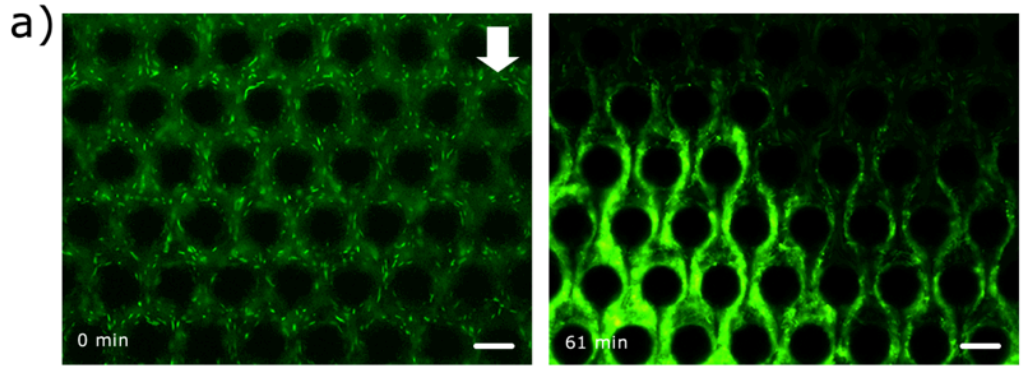

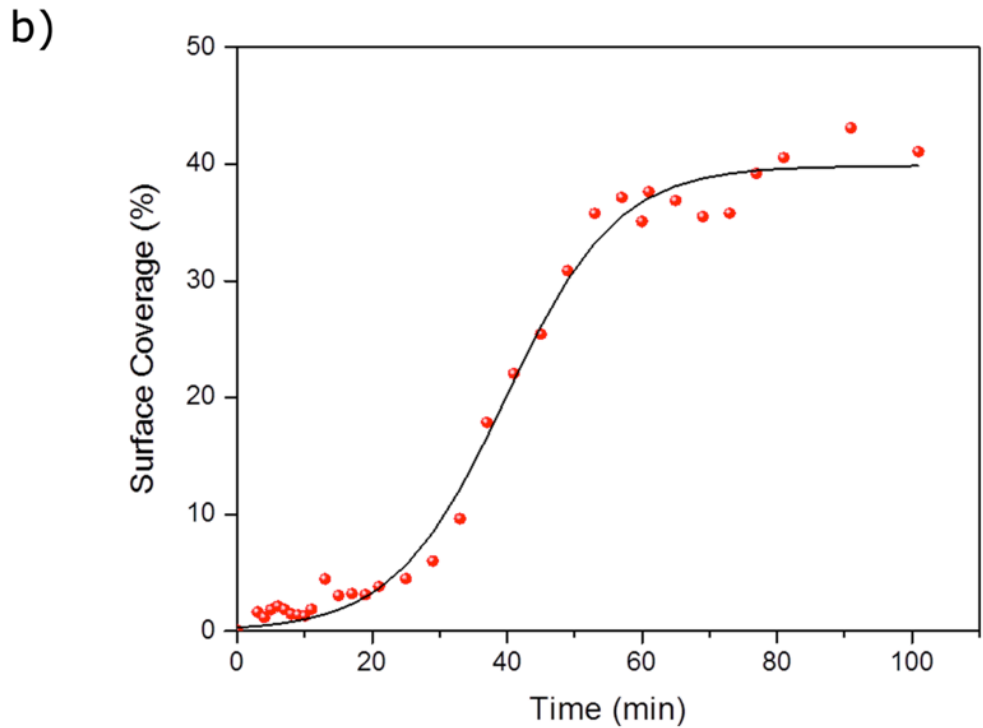

**Figure 7**: Rapid and catastrophic clogging of the channel by streamers. a) Images of the channel at two time-points. The scale bars are 50 μm. b) The graph shows a gradual increase of surface coverage from 2% (3 min) to 5% (25 min). Then, it has a dramatic surge from 5% (25 min) to 37% (57 min). After that the graph plateaus. Solid curve represents a sigmoidal curve fit to the data.

**References:**

1   Flemming, H. C. & Wingender, J. The biofilm matrix. *Nat Rev Microbiol* **8**, 623-633, doi:Doi 10.1038/Nrmicro2415 (2010).
2   Wingender, J., Neu, T. R. & Flemming, H.-C. in *Microbial extracellular polymeric substances*    1-19 (Springer, 1999).
3   Wu, C., Lim, J. Y., Fuller, G. G. & Cegelski, L. Quantitative Analysis of Amyloid-Integrated Biofilms Formed by Uropathogenic Escherichia coli at the Air-Liquid Interface. *Biophysical Journal* **103**, 464-471, doi:DOI 10.1016/j.bpj.2012.06.049 (2012).
4   Whitfield, C. Bacterial extracellular polysaccharides. *Canadian Journal of Microbiology* **34**, 415-420 (1988).
5   Flemming, H. C., Neu, T. R. & Wozniak, D. J. The EPS matrix: The "House of Biofilm cells". *J Bacteriol* **189**, 7945-7947, doi:Doi 10.1128/Jb.00858-07 (2007).
6   More, T. T., Yadav, J. S. S., Yan, S., Tyagi, R. D. & Surampalli, R. Y. Extracellular polymeric substances of bacteria and their potential environmental applications. *J Environ Manage* **144**, 1-25, doi:DOI 10.1016/j.jenvman.2014.05.010 (2014).
7   Friedman, B. A., Dugan, P. R., Pfister, R. M. & Remsen, C. C. Structure of Exocellular Polymers and Their Relationship to Bacterial Flocculation. *J Bacteriol* **98**, 1328-& (1969).
8   Drescher, K., Shen, Y., Bassler, B. L. & Stone, H. A. Biofilm streamers cause catastrophic disruption of flow with consequences for environmental and medical systems. *P Natl Acad Sci USA* **110**, 4345-4350, doi:DOI 10.1073/pnas.1300321110 (2013).
9   Karimi, A., Karig, D., Kumar, A. & Ardekani, A. Interplay of physical mechanisms and biofilm processes: review of microfluidic methods. *Lab Chip* (2015).
10  Rusconi, R., Lecuyer, S., Guglielmini, L. & Stone, H. A. Laminar flow around corners triggers the formation of biofilm streamers. *J R Soc Interface* **7**, 1293-1299, doi:DOI 10.1098/rsif.2010.0096 (2010).
11  Valiei, A., Kumar, A., Mukherjee, P. P., Liu, Y. & Thundat, T. A web of streamers: biofilm formation in a porous microfluidic device. *Lab Chip* **12**, 5133-5137, doi:Doi 10.1039/C2lc40815e (2012).
12  Stoodley, P., Cargo, R., Rupp, C. J., Wilson, S. & Klapper, I. Biofilm material properties as related to shear-induced deformation and detachment phenomena. *J. Ind. Microbiol. Biot.* **29**, 361-367 (2002).
13  Stoodley, P., Lewandowski, Z., Boyle, J. D. & Lappin-Scott, H. M. Oscillation characteristics of biofilm streamers in turbulent flowing water as related to drag and pressure drop. *Biotechnol Bioeng* **57**, 536-544 (1998).
14  Hassanpourfard, M. *et al.* Protocol for Biofilm Streamer Formation in a Microfluidic Device with Micro-pillars. *JoVE (Journal of Visualized Experiments)*, e51732-e51732 (2014).
15  Marty, A., Roques, C., Causserand, C. & Bacchin, P. Formation of bacterial streamers during filtration in microfluidic systems. *Biofouling* **28**, 551-562, doi:Doi 10.1080/08927014.2012.695351 (2012).
16  Yazdi, S. & Ardekani, A. M. Bacterial aggregation and biofilm formation in a vortical flow. *Biomicrofluidics* **6** (2012).

17  Marty, A., Causserand, C., Roques, C. & Bacchin, P. Impact of tortuous flow on bacteria streamer development in microfluidic system during filtration. *Biomicrofluidics* **8**, 014105 (2014).
18  Das, S. & Kumar, A. Formation and post-formation dynamics of bacterial biofilm streamers as highly viscous liquid jets. *Scientific reports* **4** (2014).
19  Kim, M. K., Drescher, K., Pak, O. S., Bassler, B. L. & Stone, H. A. Filaments in curved streamlines: rapid formation of Staphylococcus aureus biofilm streamers. *New J Phys* **16** (2014).
20  Hol, F. J. H. & Dekker, C. Zooming in to see the bigger picture: Microfluidic and nanofabrication tools to study bacteria. *Science* **346**, 438-+, (2014).
21  Rusconi, R., Garren, M. & Stocker, R. Microfluidics Expanding the Frontiers of Microbial Ecology. *Annu Rev Biophys* **43**, 65-91, doi:DOI 10.1146/annurev-biophys-051013-022916 (2014).
22  Mercado-Blanco, J. & Bakker, P. A. Interactions between plants and beneficial Pseudomonas spp.: exploiting bacterial traits for crop protection. *Antonie van Leeuwenhoek* **92**, 367-389, doi:10.1007/s10482-007-9167-1 (2007).
23  Chen, H.-H., Shi, J. & Chen, C.-L. Wetting dynamics of multiscaled structures. *Applied Physics Letters* **103**, 171601 (2013).
24  Arruda, E. M. & Boyce, M. C. A three-dimensional constitutive model for the large stretch behavior of rubber elastic materials. *Journal of the Mechanics and Physics of Solids* **41**, 389-412 (1993).
25  De Beer, D., Stoodley, P., Roe, F. & Lewandowski, Z. Effects of biofilm structures on oxygen distribution and mass transport. *Biotechnol Bioeng* **43**, 1131-1138 (1994).
26  De Schryver, P., Crab, R., Defoirdt, T., Boon, N. & Verstraete, W. The basics of bio-flocs technology: The added value for aquaculture. *Aquaculture* **277**, 125-137, doi:DOI 10.1016/j.aquaculture.2008.02.019 (2008).
27  Kumar, A. *et al.* Microscale confinement features can affect biofilm formation. *Microfluid Nanofluid* **14**, 895-902, doi:DOI 10.1007/s10404-012-1120-6 (2013).
28  Rusconi, R., Lecuyer, S., Autrusson, N., Guglielmini, L. & Stone, H. A. Secondary flow as a mechanism for the formation of biofilm streamers. *Biophysical journal* **100**, 1392-1399, doi:10.1016/j.bpj.2011.01.065 (2011).
29  Weaver, W. M., Milisavljevic, V., Miller, J. F. & Di Carlo, D. Fluid flow induces biofilm formation in Staphylococcus epidermidis polysaccharide intracellular adhesin-positive clinical isolates. *Appl Environ Microbiol* **78**, 5890-5896, doi:10.1128/AEM.01139-12 (2012).